\documentclass{sig-alternate}
\usepackage{lcsect}
\usepackage{csquotes}
\usepackage{amsmath}
\usepackage{epsfig}
\usepackage{epstopdf}
\usepackage{algorithm}
\usepackage{algpseudocode}
\usepackage{graphicx}
\usepackage{array}
\usepackage{float}
\usepackage{balance}
\usepackage{url}
\usepackage[caption=false]{subfig}
\DeclareCaptionType{copyrightbox}

\begin{document}

\conferenceinfo{ACM CoRR}{Computer Research Repository}
\CopyrightYear{}
\crdata{}

\title{Rapid A{\ttlit k}NN Query Processing for Fast Classification of Multidimensional Data in the Cloud}

\numberofauthors{5} 
\author{
\alignauthor
Nikolaos Nodarakis \\
  \affaddr{Computer Engineering and Informatics Department, University of Patras}\\
  \affaddr{26500 Patras, Greece}\\
  \email{nodarakis@ceid.upatras.gr}
  \alignauthor
Spyros Sioutas\\
  \affaddr{Department of Informatics, Ionian University}\\
  \affaddr{49100 Corfu, Greece}\\
  \email{sioutas@ionio.gr}
  \alignauthor
Dimitrios Tsoumakos\\
  \affaddr{Department of Informatics, Ionian University}\\
  \affaddr{49100 Corfu, Greece}\\
  \email{dtsouma@ionio.gr}
  \and
  \alignauthor
Giannis Tzimas \\
  \affaddr{Computer \& Informatics Engineering Department}\\	
  \affaddr{Technological Educational}
  \affaddr{Institute of Western Greece}  
  \affaddr{26334 Patras, Greece}\\
  \email{tzimas@cti.gr}
  \alignauthor
Evaggelia Pitoura \\
  \affaddr{Computer Science Department, University of Ioannina}\\
  \email{pitoura@cs.uoi.gr}
}

\maketitle
\begin{abstract}
A $k$-nearest neighbor ($k$NN) query determines the $k$ nearest points, using distance metrics, from a specific location. An all $k$-nearest neighbor (A$k$NN) query constitutes a variation of a $k$NN query and retrieves the $k$ nearest points for each point inside a database. Their main usage resonates in spatial databases and they consist the backbone of many location-based applications and not only (i.e. $k$NN joins in databases, classification in data mining). So, it is very crucial to develop methods that answer them efficiently. In this work, we propose a novel method for classifying multidimensional data using an A$k$NN algorithm in the MapReduce framework. Our approach exploits space decomposition techniques for processing the classification procedure in a parallel and distributed manner. To our knowledge, we are the first to study the classification of multidimensional objects under this perspective. Through an extensive experimental evaluation we prove that our solution is efficient and scalable in processing the given queries. We investigate many different perspectives that can affect the total computational cost, such as different dataset distributions, number of dimensions, growth of $k$ value and granularity of space decomposition and prove that our system is efficient, robust and scalable.
\end{abstract}


\section{Introduction}
Classification is the problem of identifying to which of a set of categories a new observation belongs, on the basis of a training set of data containing observations (or instances) whose category membership is known. One of the algorithms for data classification uses the $k$NN approach \cite{prentice} as it computes the $k$ nearest neighbors (belonging to the training dataset) of a new object and classifies it to the category that belongs the majority of its neighbors.

A $k$-nearest neighbor query \cite{rouss:kell:vinc} computes the $k$ nearest points, using distance metrics, from a specific location and is an operation that is widely used in spatial databases.  An all $k$-nearest neighbor query constitutes a variation of a $k$NN query and retrieves the $k$ nearest points for each point inside a dataset in a single query process. There is a wide diversity of applications that A$k$NN queries can be harnessed. The classification problem is one of them. Furthermore, they are widely used by location based services \cite{gkou:ver:boz}. For example, a user sends his location to a web server to process a request using a position anonymization system in order to protect his privacy from insidious acts. This anonymization system may use a $k$NN algorithm to calculate the $k$ nearest users and sends to the server their location along with the location of the user that made the request at the first place. In addition, many algorithms have been developed to optimize and speed up the join process in databases using the $k$NN approach.

Although A$k$NN is a fundamental query type, it is computationally very expensive. The naive approach is to search for every point the whole dataset in order to estimate its $k$-NN list. This leads to an $O \left(n^{2}\right)$ time complexity assuming that $n$ is the cardinality of the dataset. As a result, quite a few centralized algorithms and structures (M-trees, R-trees, space-filling curves, etc.) have been developed towards this direction \cite{ioup:shaw:sample:abdel,chen:patel,jun:mam:pap:yuf,emr:graf:krie:schu}. However, as the volume of datasets grows rapidly even these algorithms cannot cope with the computational burden produced by an A$k$NN query process. Consequently, high scalable implementations are required. Cloud computing technologies provide tools and infrastructure to create such solutions and manage the input data in a distributed way among multiple servers. The most popular and notably efficient tool is the \textit{MapReduce} \cite{dean:ghem} programming model, developed by Google, for processing large-scale data.

In this paper, we propose a method for efficient multidimensional data classification using A$k$NN queries in a single batch-based process in \textit{Hadoop} \cite{apache:hadoop, reilly}, the open source MapReduce implementation. The basic idea is to decompose the space, where the data belongs, into smaller parts, get the $k$ nearest neighbors for each point to be classified only by searching the appropriate parts and finally add it to the category it belongs based on the class that the majority of its neighbors belongs. The space decomposition relies on the data distribution of the training dataset.

More specifically, we sum up the technical contributions of our paper as follows:
\begin{itemize}
\item We present an implementation of a classification algorithm based on A$k$NN queries using MapReduce. We apply space decomposition techniques (based on data distribution) that divides the data into smaller groups and, for each point, we search for candidate $k$-NN objects only in a few groups. The granularity of the decomposition is a key factor for the performance of the algorithm and we analyze it further in Section 6.1. At first, the algorithm defines a search area for each point and investigates for $k$-NN points in the groups covered by this area. If the search area for a point does not include at least $k$ neighbors, it is gradually expanded until the desired number is reached. Finally, we classify the point to the category that belongs the majority of its neighbors. The implementation defines the MapReduce jobs with no modifications to the original Hadoop framework.

\item We provide an extension for $d > 3$ in Section 5. 

\item We evaluate our solution through an experimental evaluation against large scale data up to 3 dimensions, that studies various parameters that can affect the total computational cost of our method using real and synthetic datasets. The results prove that our solution is efficient, robust and scalable.

\end{itemize}

The rest of the paper is organized as follows: Section 2 discusses related work. Section 3 presents the initial idea of the algorithm, our technical contributions and some examples of how the algorithm works. Section 4 presents a detailed analysis of the classification process developed in Hadoop, Section 5 provides an extension for $d > 3$ and Section 6 presents the experiments that where conducted in the context of this work. Finally, Section 7 concludes the paper and Section 8 presents future steps. 
\section{Related Work}

A$k$NN queries have been extensively studied in literature. In \cite{ioup:shaw:sample:abdel}, a method based on M-trees is proposed that processes A$k$NN spatial network queries. The experimental evaluation runs over a road network dataset for small $k$ values. In addition, a structure that is popular for answering efficiently to A$k$NN queries is R-tree \cite{rouss:kell:vinc}. Pruning techniques can be combined with such structures to deliver better results \cite{chen:patel,emr:graf:krie:schu}. Mobile networks are also a domain where A$k$NN find application as shown in \cite{chatz:zeinal:lee:dik}. Their work suggest a centralized algorithm that identifies to every smartphone user its $k$ geographically nearest neighbors in $O \left(n \cdot \left(k + l\right)\right)$ time where $n$ denotes the number of users and $l$ a network-specific parameter. Moreover, efforts have been made to design low computational cost methods that execute such queries in spatial databases. For instance, \cite{yao:li:kumar} studies both the $k$NN query and the $k$NN join in a relational database and their approach guarantees to find the approximate $k$NN with only logarithmic number of page accesses in expectation with a constant approximation ratio and it could be extended to find the exact $k$NN efficiently in any fixed dimension. The works in \cite{xia:lu:chin:hu,yu:cui:wang:su} propose algorithms to answer $k$NN join.

The methods proposed above can handle data of small size in one or more dimensions, thus their use is limited in centralized environments only. During the recent years, the researchers have focused on developing approaches that are applicable in distributed environments, like our method, and can manipulate big data in an efficient manner. The MapReduce framework seems to be suitable for processing such queries. For example, in \cite{yoko:ishi:suzu} the discussed approach splits the target space in smaller cells and looks into appropriate cells where $k$-NN objects are located, but applies only in 2-dimensional data. Our method speeds up the naive solution of \cite{yoko:ishi:suzu} by eliminating the merging step, as it is a major drawback. We have to denote here that in \cite{yoko:ishi:suzu} it is claimed that the computation of the merging step can be performed in one node since we just consider statistic values. But this is not entirely true since this process can derive a notable computational burden as we increase dimensions and/or data size, something that is confirmed in the experimental evaluation. In addition, the merging step can produce sizeable groups of points, especially as $k$ increments, that can overload the first step of the A$k$NN process. Moreover, our method applies for more dimensions. Especially, for $d >= 3$ the multidimensional extension is not straightforward at all.

In \cite{stup:mich:sche}, locality sensitive hashing (LSH) is used together with a MapReduce implementation for processing $k$NN queries over large multidimensional datasets. This solution suggests an approximate algorithm like the work in \cite{zhang:li:jestes} (H-zkNNJ) but we focus on exact processing A$k$NN queries. Furthermore, A$k$NN queries are utilized along with MapReduce to speed up and optimize the join process over different datasets \cite{afra:ullm,lu:shen:chen:ooi} or support non-equi joins \cite{vern:car:li}. Moreover, \cite{bohm:krebs} makes use of a R-tree based method to process $k$NN joins efficiently. 

In \cite{cha:luo:hua:fen:fan} a minimum spanning tree based classification model is introduced and it can be viewed as an intermediate model between the traditional $k$-nearest neighbor method and cluster based classification method. Another approach presented in \cite{he:zhua:li:shi} recommends parallel implementation methods of several classification algorithms, including $k$-nearest neighbor, bayesian model, decision tree, but does not contemplate the nor the perspective of dimensionality nor parameter $k$.

In brief, our proposed method implemented in the Hadoop MapReduce framework, extends the traditional $k$NN classification algorithm and processes exact A$k$NN queries over massive multidimensional data to classify a huge amount of objects in a single batch-based process. Compared to the aforementioned solutions, our method does not focus solely on the join operator but provides a more generalized framework to process A$k$NN queries. The experimental evaluation considers a wide diversity of factors that can affect the execution time such as the value of $k$, the granularity of space decomposition, dimensionality and data distribution.

\section{Overview of Classification Algorithm}

In this section, we first define some notation and provide some definitions used throughout this paper. Table \ref{tab1} lists the symbols and their meanings. Next, we give a brief review of the method our solution relies on and then we extend it for more dimensions and tackle some performance issues.

\begin{table}
\centering
\caption{Symbols and their meanings}
\label{tab1}
\begin{tabular}{|l|l|} \hline
$n$ & granularity of space decomposition \\ \hline
$k$ & number of nearest neighbors \\ \hline
$d$ & dimensionality \\ \hline
$D$ & a $d$-dimensional metric space \\ \hline
$dist(r,s)$ & the distance from $r$ to $s$ \\ \hline
$kNN(r,S)$ & the $k$ nearest neighbors of $r$ from $S$ \\ \hline
$AkNNC(R,S)$ & $\forall r \in R$ classify $r$ based on $kNN(r,S)$ \\ \hline
$ICCH$ & interval, cell cube or hypercube \\ \hline
$ICSH$ & interval, circle, sphere or hypersphere\\ \hline
$I$ & input dataset \\ \hline
$T$ & training dataset \\ \hline
$c_r$ & the class of point $r$ \\ \hline
$C_T$ & the set of classes of dataset $T$ \\ \hline
$S_I$ & size of input dataset \\ \hline
$S_T$ & size of training dataset \\ \hline
$M$ & total number of Map tasks \\ \hline
$R$ & total number of Reduce tasks \\ \hline
\end{tabular}
\end{table}

\subsection{Definitions}

We consider points in a $d$-dimensional metric space $D$. Given two points $r$ and $s$ we define as  $dist(r,s)$ the distance between $r$ and $s$ in $D$. In this paper, we used the distance measure of Euclidean distance

\begin{displaymath}dist(r,s) = \sqrt{\sum\nolimits_{i=1}^d \left(r[i] - s[i] \right)^2} \end{displaymath}

\noindent
where $r[i]$ (respectively $s[i]$) denote the value of $r$ (respectively $s$) along the $i^{th}$ dimension in $D$. Without loss of generality, alternative distance measures (i.e. Manhattan distance) can be applied to our solution.

\newdef{definition}{Definition}
\begin{definition}
\textbf{$k$NN:} Given a point $r$, a dataset $S$ and an integer $k$, the $k$ nearest neighbors of $r$ from $S$, denoted as $kNN(r,S)$, is a set of $k$ points from $S$ such that $\forall p \in kNN(r,S)$, $\forall q \in \{S - kNN(r,S)\}, dist(p,r) < dist(q, r)$.
\end{definition}

\begin{definition}
\textbf{A$k$NN:} Given two datasets $R, S$ and an integer $k$, the all $k$ nearest neighbors of $R$ from $S$, named $AkNN(R,S)$, is a set of pairs $(r,s)$ such that $AkNN(R,S) = \{\left(r,s\right): \forall r \in R, \forall s \in kNN(r,S)\}$.
\end{definition}

\begin{definition}
\textbf{A$k$NN Classification:} Given two datasets $R, S$ and a set of classes $C_S$ where points of $S$ belong, the classification process produces a set of pairs $(r, c_r)$, denoted as $AkNNC(R,S)$, such that $AkNNC(R,S) = \{\left(r, c_r\right): \forall r \in R, c_r \in C_S\}$ where $c_r$ is the class where the majority of $kNN(r,S)$ belong $\forall r \in R$.
\end{definition}

\subsection{Classification Using Space Decomposition}

Consider a training dataset $T$, an input dataset $I$ and a set of classes $C_T$ where points of $T$ belong. First of all, we define as \textit{target space} the space enclosing the points of $I$ and $T$. The parts that occur when we decompose the target space for 1-dimensional objects are called \textit{intervals}. Respectively, we call \textit{cells} and \textit{cubes} the parts in case of 2 and 3-dimensional objects and hypercubes for $d > 3$. For a new $1D$ point $p$, we define as \textit{boundary interval} an interval centred at $p$ that covers $k$-NN elements. Respectively, we define the \textit{boundary circle} and \textit{boundary sphere} for $2D$ and $3D$ points and the \textit{boundary hypersphere} for $d > 3$. The notion of hypercube and hypersphere are analyzed further in Section 5. When the boundary ICSH centred in an ICCH $icch_1$, intersects the bounds of an other $icch_2$ we say an \textit{overlap} occurs on $icch_2$. Finally, for a point $i \in I$, we define as \textit{updates} of $kNN(i,T)$ the existence of many different instances of $kNN(i,T)$ that need to be unified to a final set.

We place the objects of $T$ on the target space according to their coordinates. The main idea of equal-sized space decomposition is to partition the target space into $dn$ equal sized ICCHs where $n$ and the size of each ICCH are user defined. Each ICCH contains a number of points of $T$. Moreover, we define a new layer over the target space according to $C_T$ and $\forall t \in T, c_t \in C_T$. In order to estimate $AkNNC(I,T)$, we investigate $\forall i \in I$ for $k$-nearest neighbors only in a few ICCHs, thus bounding the number of computations that need to be performed for each $i$. 

\subsection{Previous Work}
A very preliminary study of naive A$k$NN solutions is presented in \cite{yoko:ishi:suzu} and uses a simple cell decomposition technique to process A$k$NN queries on two different datasets, i.e. $I$ and $T$. The objects consisting both datasets are $2$-dimensional points having only one attribute, the coordinate vector and the target space comprises of $2^{n} \times 2^{n}$ equal-sized cells. 

The elements of both datasets are placed on the target space according to their coordinate vector and a cell decomposition is applied. For a point $i \in I$ it is expected that its $kNN(i,T)$ will be located in a close range area defined by nearby cells. At first, we look for candidate $k$-NN points inside the cell that $i$ belongs in the first place, name it $cl$. If we find at least $k$ elements we draw the boundary circle. There is a chance the boundary circle centred at $cl$ overlaps some neighboring cells. In this case, we need to investigate for possible $k$-NN objects inside these overlapped cells in order to create the final $k$-NN list. If no overlap occurs, the $k$-NN list of $i$ is complete. Next, we present an example for better understanding of the algorithm.

\begin{figure}[t]
\centering
\epsfig{file=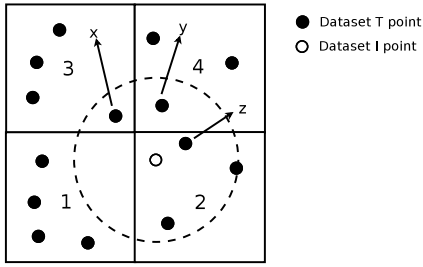,scale=.52}
\caption{$k$NN process using cell decomposition ($k$ = 3)}
\label{fig1}
\end{figure}

Figure \ref{fig1} illustrates an example of the A$k$NN process of a point in dataset $I$ using a query for $k = 3$. Initially, the point looks for $k$-NN objects inside cell 2. Since there exist at least 3 points of dataset $T$ in cell 2 the boundary circle can be drawn. The boundary circle overlaps cells 1,3 and 4, so we need to investigate for additional $k$-NN objects inside them. The algorithm outputs an instance of the $k$-NN list for every overlapped cell. These instances need to be unified into a $k$-NN list containing the final points ($x, y$, $z$).

\begin{figure}[t]
\centering
\epsfig{file=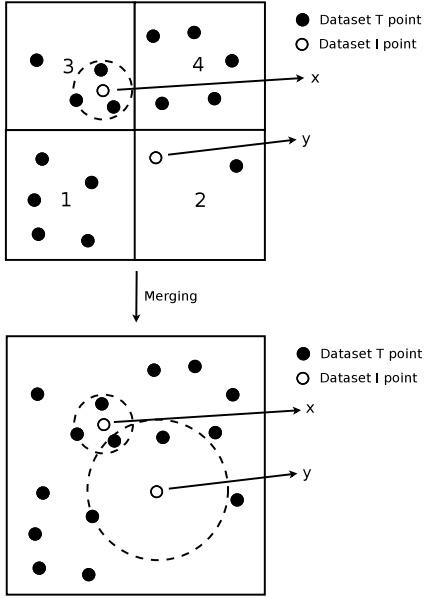,scale=.52}
\caption{Issue of the merging step before the $k$NN process ($k = 3$)}
\label{fig2}
\end{figure}

This approach, as described above, fails to draw the boundary circle if $cl$ contains less than $k$ points. The solution to the problem is simple. At first, we check the number of points that fall into every cell. If we find a cell with less than $k$ points we merge it with the neighboring cells to assure that it will contain the required number of objects. The way the merging step is performed relies on the principles of hierarchical space decomposition used in quad-trees \cite{samet}. Note that this is the reason why the space decomposition involves $2^{n} \times 2^{n}$ cells. This imposes two more steps that need to be done before we begin calculating $kNN(i,T)$. In the beginning, a counting phase needs to be performed followed by a merging step in order to overcome the issue mentioned above. This preprocessing phase induces additional cost to the total computation and, as shown in the experiments, the merging step can lead to a bad algorithmic behavior.

\begin{figure}
\centering
\epsfig{file=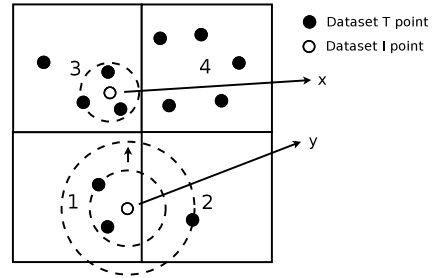,scale=.52}
\caption{Increasing the search range until it covers at least $k$ neighbors ($k = 3$)}
\label{fig3}
\end{figure}

\subsection{Technical Contributions}

In this subsection, we extend the previous method for more dimensions and adapt it to the needs of the classification problem. Moreover, we analyze some drawbacks of the method studied in \cite{yoko:ishi:suzu} and propose a mechanism to make the algorithm more efficient.

Firstly, we have a training dataset $T$, an input dataset $I$ and a set of classes $C_T$ where points of $T$ belong. The only difference now is that the points in the training dataset have one more attribute, the class they belong. In order to compute $AkNNC(I,T)$, a classification step is executed after the construction of the $k$-NN lists. The class of every new object is chosen based on the class membership of its $k$-nearest neighbors. Furthermore, we extend the solution presented in \cite{yoko:ishi:suzu} for more dimensions, and now the space is decomposed in $2^{dn}$ ICCHs.

\begin{figure*}
\centering
\resizebox{\linewidth}{8cm}{\epsfig{file=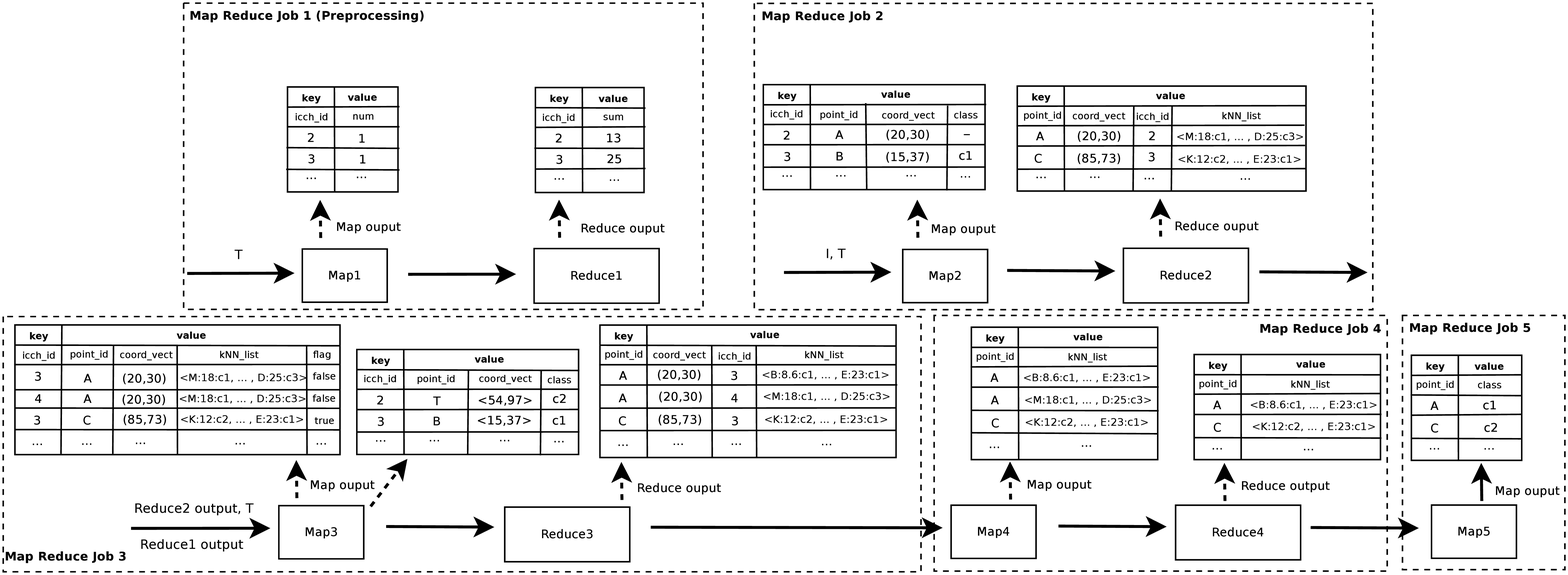}}
\caption{Overview of the A$k$NN classification process}
\label{fig4}
\end{figure*}

As mentioned before, the simple solution presented in \cite{yoko:ishi:suzu} has one major drawback which is the merging step. Figure \ref{fig2} depicts a situation where the merging step of the original method can significantly increase the total cost of the algorithm. Consider two points $x$ and $y$ entering cells 3 and 2 respectively and $k = 3$. We can draw point's $x$ boundary circle since cell 3 includes at least $k$ elements. On the contrary, we cannot draw the boundary circle of point $y$, so we need to unify cells 1 through 4 into one bigger cell. Now point $y$ can draw its boundary circle but we overload point's $x$ $k$-NN list construction with redundant computations. In the first place, the $k$-NN list of point $x$ would only need 4 distance calculations to be formed. After the merging step we need to perform 15, namely almost 4 times more than before and this would happen for all points that would join cells 1,3 and 4 in the first place.

In order to avoid a scenario like above, we introduce a mechanism where only points that cannot find at least $k$-nearest neighbors in the ICCH in the first place proceed to further actions. Let a point $p$ joining an ICCH $icch$ that encloses $l$ neighbors and $l < k$. Instead of performing a merging step, we draw the boundary ICSH based on these $l$ neighbors. Then, we check if the boundary ICSH overlaps any neighboring ICCHs. In case it does, we investigate if the boundary ICSH covers at least $k$ elements in total. In case it does, then we are able to build the final $k$-NN list of the point by unifying the individual $k$-NN lists that are derived for every overlapped ICCH. In case the boundary ICSH does not cover at least $k$ objects in total or does not overlap any ICCHs then we gradually increase its search range until the prerequisites are fulfilled.

Figure \ref{fig3} explains this issue. Consider two points $x$ and $y$ entering cells 3 and 1 respectively and $k = 3$. We observe that cell 3 contains 4 neighbors and point $x$ can draw its boundary circle that covers $k$-NN elements. However, the boundary circle centred at point $y$ does not cover $k$-NN elements in the first place. Consequently, we gradually increase its search range until the boundary circle encloses at least $k$-NN points. Note that eliminating the merging step, we also relax the condition of decomposing the target space into $2^{dn}$ equal-sized splits and generalize it to $dn$ equal-sized splits.

Summing up, our solution can be implemented as a series of MapReduce jobs as shown below. These MapReduce jobs will be analyzed in detail in Section 4:

\begin{enumerate}
  \item \textbf{Distribution Information}. Count the number of points of $T$ that fall into each ICCH.
  \item \textbf{Primitive Computation Phase}. Calculate possible $k$-NN points $\forall i \in I$ from $T$ in the same ICCH.
  \item \textbf{Update Lists}. Draw the boundary ICSH $\forall i \in I$ and increase it, if needed, until it covers at least $k$-NN points of $T$. Check for overlaps of neighboring ICCHs and derive updates of $k$-NN lists.
  \item \textbf{Unify Lists}. Unify the updates of every $k$-NN list into one final $k$-NN list $\forall i \in I$.
  \item \textbf{Classification}. Classify all points of $I$.
  
\end{enumerate}

In Figure \ref{fig4}, we illustrate the working flow of the A$k$NN classification process which consists of 5 MapReduce jobs. Each MapReduce job is studied in detail in the next section. Note, that the first MapReduce job acts as a preprocessing step and its results are provided as additional input in MapReduce Job 3 and that the preprocessing step is executed only once for $T$. 

\section{Detailed Analysis of Classification Procedure}
In this section, we present a detailed description of the classification process as implemented in the Hadoop framework. The whole process consists of five MapReduce jobs which are divided into three phases. Phase one estimates the distribution of $T$ over the target space. Phase two determines $kNN(i,T), \forall i \in I$ and phase three estimates $AkNNC(I,T)$. The records in $T$ have the format $<$point\textunderscore id, coordinate\textunderscore vector, class$>$ and in $I$ have the format $<$point\textunderscore{id}, coordinate\textunderscore{vector}$>$. Furthermore, parameters $n$ and $k$ are defined by the user. In the following subsections, we describe each MapReduce job separately and analyze the Map and Reduce functions that take place in each one of them. For each MapReduce job, we also quote pseudo-code for better understanding of the Map and Reduce functions and proceed to time and space complexity analysis.

\subsection{Getting Distribution Information of Training Dataset}
This MapReduce job is a preprocessing step required by subsequent MapReduce jobs that receive its output as additional data. In this step, we decompose the entire target space and count the number of points of $T$ that fall in each ICCH. Below, we sum up the Map and Reduce functions consisting this MapReduce process.

\begin{algorithm}
\floatname{algorithm}{MapReduce Job}
\newcommand{\algorithmname}{MapReduce Job}
\renewcommand{\thealgorithm}{1}
\caption{\null}
\begin{algorithmic}[1]
\Function{Map}{$k1, v1$}
\State $coord = \textbf{getCoord}(v1);$ $icch\textunderscore{id} = \textbf{getId}(coord)$
\State $\textbf{output}(icch\textunderscore{id}, 1);$
\EndFunction
\Statex
\Function{Reduce}{$k2, v2$}
\State $sum = 0;$
\ForAll {$v \in v2$}
\State {$sum = sum + \textbf{getSum}(v);$}
\State $\textbf{output}(k2, sum);$
\EndFor
\EndFunction
\end{algorithmic}
\end{algorithm}

The \textit{Map} function takes as input records with the training dataset format, estimates the ICCH id for each point based on its coordinates and outputs a key-value pair where the key is ICCH id and the value is number 1. The \textit{Reduce} function receives the key-value pairs from the Map function and for each ICCH id it outputs the number of points of $T$ that belong to it.  

Each Map task needs $O \left(S_T / M\right)$ time to run. Each Reduce task needs $O \left(dn / R\right)$ time to run as the total number of ICCHs is $dn$. So, the size of the output will be $O \left(2dn \cdot c_{si}\right)$, where $c_{si}$ is the size of sum and icch\textunderscore{id} for an output record.

\subsection{Estimating Primitive Phase Neighbors of A{\subsecit k}NN Query}
In this stage, we concentrate all training ($L_T$) and input ($L_I$) records for each ICCH and compute possible $k$-NN points for each item in $L_I$ from $L_T$ inside the ICCH. Below, we condense the Map and Reduce functions. We use two Map functions, one for each dataset, as seen in MapReduce Job 2 pseudo-code. 

\begin{algorithm}
\floatname{algorithm}{MapReduce Job}
\newcommand{\algorithmname}{MapReduce Job}
\renewcommand{\thealgorithm}{2}
\caption{\null}
\begin{algorithmic}[1]
\Function{Map1}{$k1, v1$}
\State $coord = \textbf{getCoord}(v1);$ $p\textunderscore{id} = \textbf{getPointId}(v1);$
\State $class = \textbf{getClass}(v1);$ $icch\textunderscore{id} = \textbf{getId}(coord);$
\State $\textbf{output}(icch\textunderscore{id}, <p\textunderscore{id}, coord, class>)$;
\EndFunction
\Statex
\Function{Map2}{$k1, v1$}
\State $coord = \textbf{getCoord}(v1);$ $p\textunderscore{id} = \textbf{getPointId}(v1);$
\State $icch\textunderscore{id} = \textbf{getId}(coord)$;
\State $\textbf{output}(icch\textunderscore{id}, <p\textunderscore{id}, coord>)$;
\EndFunction
\Statex
\Function{Reduce}{$k2, v2$}
\State $L_T = \textbf{getTrainingPoints}(v2);$
\State $L_I = \textbf{getInputPoints}(v2);$
\ForAll {$p \in L_I$}
\State $L = List\{\};$
\ForAll {$t \in L_T$}
\State $L.add(\textbf{new} Record(t, dist(p, t), t.class));$
\EndFor
\State $\textbf{output}(p.id, <p.coord, k2, \textbf{getKNN}(L)>);$
\EndFor
\EndFunction
\end{algorithmic}
\end{algorithm}

For each point $t \in T$, \textit{Map1} outputs a new key-value pair in which the ICCH id where $t$ belongs is the key and the value consists of the id, coordinate vector and class of $t$. Similarly, for each point $i \in I$, \textit{Map2} outputs a new key-value pair in which the ICCH id where $i$ belongs is the key and the value consists of the id and coordinate vector of $i$. The \textit{Reduce} function receives a set of records from both Map functions with the same ICCH ids and separates points of $T$ from points of $I$ into two lists, $L_T$ and $L_I$ respectively. Then, the Reduce function calculates the distance for each point in $L_I$ from $L_T$, estimates the $k$-NN points and forms a list $L$ with the format $<p_1,d_1,c_1\mathrm{:\ldots:}p_k,d_k,c_k>$, where $p_i$ is the $i$-th NN point, $d_i$ is its distance and $c_i$ is its class. Finally, for each $p \in L_I$, \textit{Reduce} outputs a new key-value pair in which the key is the id of $p$ and the values comprises of the coordinate vector, ICCH id and list $L$ of $p$. 

Each Map1 task needs $O \left(S_T / M\right)$ time and each Map2 task needs $O \left(S_I / M\right)$ time to run. For a Reduce task, suppose $u_i$ and $t_i$ the number of input and training points that are enclosed in an ICCH in the $i$-th execution of a Reduce function and $1 \leq i \leq dn / R$. The Reduce task needs $O \left(\sum\nolimits_{i} u_i \cdot t_i\right)$. Let $L_T$ to be the size of $k$-NN list and icch\textunderscore{id} $\forall i \in I$. The output size is $O \left(S_I \cdot L_T\right)$, which is $O\left(S_I\right)$.

\subsection{Checking for Overlaps and Updating {\subsecit k}-NN Lists}
In this step, at first we gradually increase the boundary ICSH, where necessary, until it includes at least $k$ points. Then, we check for overlaps between neighboring ICCHs and derive updates of the $k$-NN lists. The Map and Reduce functions are outlined in MapReduce Job 3 pseudo-code.

\begin{algorithm}
\floatname{algorithm}{MapReduce Job}
\newcommand{\algorithmname}{MapReduce Job}
\renewcommand{\thealgorithm}{3}
\caption{\null}
\begin{algorithmic}[1]
\Function{Map1}{$k1, v1$}
\State Same as Map1 function from MapReduce Job 2
\EndFunction
\Statex
\Function{Map2}{$k1, v1$}
\State $c = \textbf{getCoord}(v1);$ $p\textunderscore{id} = \textbf{getPointId}(v1);$
\State $kNN = \textbf{getKNNList}(v1);$ $r = \textbf{getRadius}(kNN);$
\While {$kNN.size() < k$}
\State $\textbf{increase}(r);$
\State $kNN.addAll(\textbf{getNeighbors}(r));$
\EndWhile
\State $oICCHs = \textbf{getOverlappedICCHs}(r);$
\If {$oICCHs.size() > 0$}
\ForAll {$icch \in oICCHs$}
\State $\textbf{output}(icch, <p\textunderscore{id}, c, kNN, false>);$
\EndFor
\Else 
\State {$\textbf{output}(\textbf{getId}(c), <p\textunderscore{id}, c, kNN, true>);$}
\EndIf
\EndFunction
\Statex
\Function{Reduce}{$k2, v2$}
\State $L_T = \textbf{getTrainingPoints}(v2);$
\State $L_I = \textbf{getInputPoints}(v2);$
\ForAll {$p \in I$}
\If {$p.flag == true$}
\State $\textbf{output}(p.id, <p.coord, key, p.kNN>);$
\Else
\State $L = List\{\};$
\ForAll {$t \in T$}
\State $L.add(\textbf{new} Record(t, dist(p, t), t.class));$
\EndFor
\State $L_f = \textbf{finalKNN}(L, p.kNN);$
\State $\textbf{output}(p.id, <p.coord, key, L_f>);$
\EndIf
\EndFor
\EndFunction
\end{algorithmic}
\end{algorithm}

The \textit{Map1} function is exactly the same as \textit{Map1} function in the previous job. For each point $i \in I$, function \textit{Map2} computes the overlaps between neighboring ICCHs. If no overlap occurs, it does not need to perform any additional steps and outputs a key-value pair in which ICCH id is the key and the value consists of id, coordinate vector and list $L$ of $i$ and a flag \textit{true} which implies that no further process is required. Otherwise, for every overlapped ICCH it outputs a new record where ICCH id$'$  (id of an overlapped ICCH) is the key and the value consists of id, coordinate vector and list $L$ of $i$ and a flag \textit{false} that indicates we need to search for possible $k$-NN objects inside the overlapped ICCHs. The \textit{Reduce} function receives a set of points with the same ICCH ids and separates the points of $T$ from points of $I$ into two lists, $L_T$ and $L_I$ respectively. After that, the Reduce function performs extra distance calculations using the points in $L_T$ and updates $k$-NN lists for the records in $L_I$. Finally, for each $p \in L_I$ it generates a record 
in which the key is the id of $p$ and the values comprises of the coordinate vector, ICCH id and list $L$ of $p$.

Each Map1 task needs $O \left(S_T / M\right)$ time to run. Consider an unclassified point $p$ initially belonging to an ICCH $icch$. Let $r$ be the number of times we increase the search range for $p$ and $icchov$ the number of ICCHs that may be overlapped for $p$. For each Map2 task the $i$-th execution of the Map function performs $icchov_i + r_i$ steps, where $1 \leq i \leq S_I / M$. So, each Map2 task runs in $O \left(\sum\nolimits_{i} icchov_i + r_i\right)$ time. For a Reduce task, suppose $u_i$ and $t_i$ the number of points of $I$ and $T$ respectively that are enclosed in an ICCH in the $i$-th execution of a Reduce function and $1 \leq i \leq dn / R$. The Reduce task needs $O \left(\sum\nolimits_{i} u_i \cdot t_i\right)$. The size of updated records is a fraction of $S_I$. So, the size of the output is also $O \left(S_I\right)$.

\subsection{Unifying Multiple {\subsecit k}-NN Lists}
During the previous step it is possible that multiple updates of a point's $k$-NN list might occur. This MapReduce job tackles this problem and unifies possible multiple lists into one final $k$-NN list for each point $i \in I$. The Map and Reduce functions are summarized at MapReduce Job 4 pseudo-code below.

\begin{algorithm}
\floatname{algorithm}{MapReduce Job}
\newcommand{\algorithmname}{MapReduce Job}
\renewcommand{\thealgorithm}{4}
\caption{\null}
\begin{algorithmic}[1]
\Function{Map}{$k1, v1$}
\State $\textbf{output}(\textbf{getPointId}(v1), \textbf{getKKN}(v1));$
\EndFunction
\Statex
\Function{Reduce}{$k2, v2$}
\State $L = List\{\};$
\ForAll {$v \in v2$}
\State $L.add(v);$
\EndFor
\State $\textbf{output}(k2, \textbf{unifyLists}(L));$
\EndFunction
\end{algorithmic}
\end{algorithm}

The \textit{Map} function receives the records of the previous step and extracts the $k$-NN list for each point. For each point $i \in I$, it outputs a key-value pair in which the key is the id of $i$ and the value is the list $L$. The \textit{Reduce} function receives as input key-value pairs with the same key and computes $kNN(i,T), \forall i \in I$. The key of an output record is again the id of $i$ and the value consists of $kNN(i,T)$.

Each Map task runs in $O \left(S_I / M\right)$. For each Reduce task, assume $updates_i$ the number of updates for the $k$-NN list of an unclassified point in the $i$-th execution of a Reduce function, where $1 \leq i \leq |N_I| / R$ and $|N_I|$ the number of points in input dataset. Then, each Reduce task needs $O \left(\sum\nolimits_{i} updates_i\right)$ to run. Let, $I_{id}$ the size of ids of all points in $I$ and $L_{final}$ is the size of the final $k$-NN list $\forall i \in I$. The size of $L_{final}$ is constant and $I_{id}$ is $O\left(S_I\right)$. Consequently, the size of the output is $O\left(S_I\right)$.

\subsection{Classifying Points}
This is the final job of the whole classification process. It is a Map-only job that classifies the input points based on the class membership of their $k$-NN points. The Map function receives as input records from the previous job and outputs $AkNNC(I,T)$. MapReduce Job 5 pseudo-code depicts the functionality of this job. 

\begin{algorithm}
\floatname{algorithm}{MapReduce Job}
\newcommand{\algorithmname}{MapReduce Job}
\renewcommand{\thealgorithm}{5}
\caption{\null}
\begin{algorithmic}[1]
\Function{Map}{$k1, v1$}
\State $H = HashMap<Class, Occurences>\{\};$
\State $H = \textbf{findClassOccur}(v1);$
\State $max = 0; maxClass = \textbf{null};$
\ForAll {$entry \in H$}
\If {$entry.occur > max$}
\State $max = entry.occur;$
\State $maxClass = entry.class;$
\EndIf
\EndFor
\State $\textbf{output}(\textbf{getPointId}(v1), maxClass);$
\EndFunction
\end{algorithmic}
\end{algorithm}

Each Map task runs in $O \left(S_I / M\right)$ time and the size of the output is $O \left( S_I \right)$.

\lcsection{EXTENSION FOR {\secit d} > 3}

Here we provide the extension of our method for $d > 3$. In geometry, a hypercube is a $n$-dimensional analogue of a square ($n = 2$) and a cube ($n = 3$) and is also called a n-cube (i.e. 0-cube is a hypercube of dimension zero and represents a point). It is a closed, compact and convex figure that consists of groups of opposite parallel line segments aligned in each of the space's dimensions, perpendicular to each other and of the same length.

Respectively, an $n$-sphere is a generalization of the surface of an ordinary sphere to a $n$-dimensional space. Spheres of dimension $n > 2$ are called hyperspheres. For any natural number $n$, an $n$-sphere of radius $r$ is defined as a set of points in $(n + 1)$-dimensional Euclidean space which are at distance $r$ from a central point and $r$ may be any positive real number. So, the $n$-sphere centred at the origin is defined by: 
\begin{displaymath}S^{n} = \{x \in \Re^{n + 1}: \parallel x \parallel = r \}\end{displaymath}

\begin{figure}
\centering
\epsfig{file=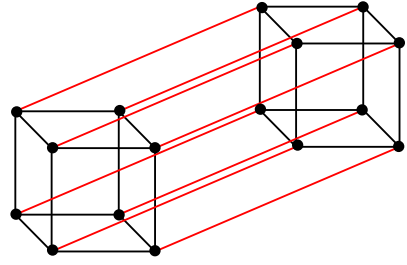,scale=.50}
\caption{Creating a 4-cube from a 3-cube}
\label{fig5}
\end{figure}

Figure \ref{fig5} displays how to create a hypercube for $d = 4$ (4-cube) from a cube for $d = 3$. Regarding our solution for $d > 3$, the target space now is decomposed into equal-sized $d$-dimensional hypercubes and in the first place we investigate for $k$-NN points in each hypercube. Next, we draw the boundary hypersphere and increase it, if needed, until it bounds at least $k$ neighbors. Finally, we inspect for any overlaps between the boundary hypersphere and neighboring hypercubes, we build the final $k$-NN list for each unclassified point and categorize it according to class majority of its $k$-NN list.

\section{Experimental Evaluation}

In this section, we conduct a series of experiments to evaluate the performance of our method under many different perspectives such as the value of $k$, the granularity of space decomposition, dimensionality and data distribution.

Our cluster includes 32 computing nodes (VMs), each one of which has four 2.1 GHz CPU processors, 4 GB of memory, 40 GB hard disk and the nodes are connected by 1 gigabit Ethernet. On each node, we install Ubuntu 12.04 operating system, Java 1.7.0\textunderscore{40} with a 64-bit Server VM, and Hadoop 1.0.4. To adapt the Hadoop environment to our application, we apply the following changes to the default Hadoop configurations: the replication factor is set to 1; the maximum number of Map and Reduce tasks in each node is set to 3, the DFS chunk size is 256 MB and the size of virtual memory for each Map and Reduce task is set to 512 MB.

We evaluate the following approaches in the experiments: 

\begin{itemize}
\item kdANN is the solution proposed in \cite{yoko:ishi:suzu} along with the extension (which invented and implemented by us) for more dimensions, as described in Section 3, in order to be able to compare it with our solution.

\item kdANN+ is our solution for $d$-dimensional points without the merging step as described in Section 3.
\end{itemize}

We evaluate our solution using both real\footnote{The real dataset is part of the Canadian Planetary Emulation Terrain 3D Mapping Dataset, which is a collection of 3-dimensional laser scans gathered at two unique planetary analogue rover test facilities in Canada. The dataset provides the coordinates $(x,y,z)$ for each laser scan in meters. \url{http://asrl.utias.utoronto.ca/datasets/3dmap/}} and synthetic datasets. We create $1D$ and $2D$ datasets from the real dataset keeping the $x$ and the $(x,y)$ coordinates respectively. We process the dataset to fit into our solution  (i.e. normalization) and we end up with $1D$, $2D$ and $3D$ datasets that consist of approximately 19,000,000 points and follow a power law like distribution. From each dataset, we extract a fraction of points (10\%) that are used as a training dataset. Respectively, we create 1, 2 and 3-dimensional datasets with uniformly distributed points, each dataset has 19,000,000 points and the training datasets contain 1,900,000 points. For each point in a training dataset we assign a class based on its coordinate vector. The file sizes of datasets are: 

\begin{enumerate}
  \item Real Dataset
  \begin{enumerate}
    \item $1D$: Input set size is 309.5 MB and training set size is 35 MB
    \item $2D$: Input set size is 403.5 MB and training set size is 44.2 MB
    \item $3D$: Input set size is 523.7 MB and training set size is 56.2 MB
  \end{enumerate}
  \item Synthetic Dataset
  \begin{enumerate}
    \item $1D$: Input set size is 300.7 MB and training set size is 33.9 MB
    \item $2D$: Input set size is 359.2 MB and training set size is 39.8 MB
    \item $3D$: Input set size is 478.5 MB and training set size is 51.7 MB
  \end{enumerate}
\end{enumerate}

\subsection{Tuning parameter {\subsecit n}}

One major aspect in the performance of the algorithm is the tuning of granularity parameter $n$. In this experiment, we explain how to select a value of $n$ in order to succeed in achieving the shortest execution time. Each time the target space is decomposed into $2^{dn}$ equal parts in order for kdANN to be able to perform the merging step, as described in Section 3.

In the case of power law distributions, we choose higher values of $n$ compared to uniform distributions since we want to discretize the target space into splits that contain as fewer points as possible in order to avoid an overload of the primitive computation phase. On the other hand, as $n$ increases, the number of update steps also increases and this can overwhelm the A$k$NN process if the number of derived instances of the $k$-NN lists is massive. Regarding uniform distributions, we wish to create larger splits, but again not too big, in order to avoid executing many update steps. Each time, the selection of $n$ depends on the infrastructure of the cluster.

\begin{figure}
\centering
\epsfig{file=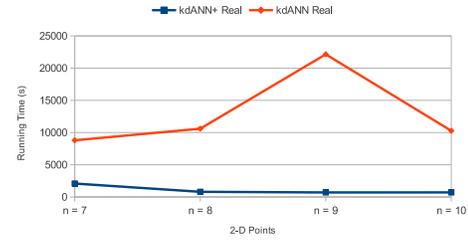,scale=.40}
\caption{Effect of $n$ (Real Dataset $2D$)}
\label{fig6}
\end{figure}

In Figure \ref{fig6}, we depict how execution time varies as we alter value $n$ in case of 2-dimensional real dataset for $k = 5$. In case of kdANN+, we notice that as parameter $n$ grows the execution time drops and achieves its lowest value for $n = 9$ and slightly increases for $n = 10$. In contrary, the execution time for kdANN increases until $n = 9$ and drops significantly for $n = 10$. Moreover, its lowest achieved value is almost ten times bigger than kdANN+. Considering the above, we deduce that for power law distributions kdANN+ outperforms kdANN as $n$ changes. In addition, we conclude that the merging step affects greatly the performance of kdANN and creates a wide divergence in total running time as $n$ mutates.

\begin{figure}
\centering
\epsfig{file=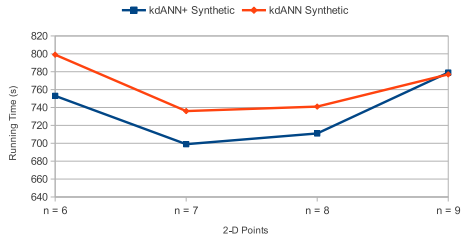,scale=.40}
\caption{Effect of $n$ (Synthetic Dataset $2D$)}
\label{fig7}
\end{figure}

Figure \ref{fig7}, presents the results of execution time for both methods when datasets follow a uniform distribution. Again, kdANN+ performs better than kdANN but now the curve of execution time presents a same behavior for both methods and when $n = 7$ the minimum running time is achieved. Observing the exported results from Figures \ref{fig6} and \ref{fig7}, we confirm our claim that we choose higher values of $n$ in case of power law distribution datasets, compared to uniformly distributed datasets, in order to minimize the total execution time.

The results for $1D$ and $3D$ points follow the same trend. In the case of real datasets, we pick value $n$ that maximizes the performance of kdANN+ since kdANN presents a bad algorithmic behavior regardless of value $n$, as shown in the majority of experiments that follow. So, for the rest of our experiments we set the value $n$ as summarized below: 

\begin{enumerate}
  \item Real Dataset
  \begin{enumerate}
    \item $1D$: $n = 18$
    \item $2D$: $n = 9$
    \item $3D$: $n = 7$
  \end{enumerate}
  \item Synthetic Dataset
  \begin{enumerate}
    \item $1D$: $n = 16$
    \item $2D$: $n = 7$
    \item $3D$: $n = 5$
  \end{enumerate}
\end{enumerate}

\subsection{Effect of {\subsecit k} and Effect of Dimensionality}

In this experiment, we evaluate both methods using real and synthetic datasets and record the execution time as $k$ increases for each dimension. Finally, we study the effect of dimensionality on the performance of kdANN and kdANN+.

\subsubsection{Effect of $k$ for different dimensions}
\begin{figure}[t]
\centering
\epsfig{file=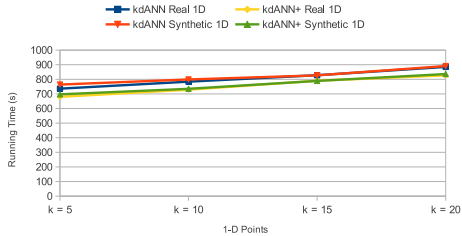,scale=.40}
\caption{Effect of $k$ for $d = 1$}
\label{fig8}
\end{figure}

Figure \ref{fig8} presents the results for kdANN and kdANN+ by varying $k$ from 5 to 20 on real and synthetic datasets. In terms of running time, kdANN+ always perform best, followed by kdANN and each method behave in the same way for both datasets, real and synthetic. As the value of $k$ grows, the size of each intermediate record becomes larger respectively. Consequently, the data processing time increments. Moreover, as the number of neighbors we need to estimate each time augments, we need to search into more intervals for possible $k$-NN points as the boundary interval grows larger.

\begin{figure}[t]
\centering
\epsfig{file=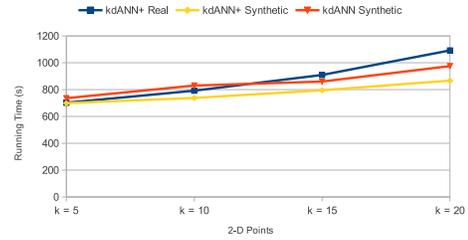,scale=.40}
\caption{Effect of $k$ for $d = 2$}
\label{fig9}
\end{figure}

\begin{figure}[t]
\centering
\epsfig{file=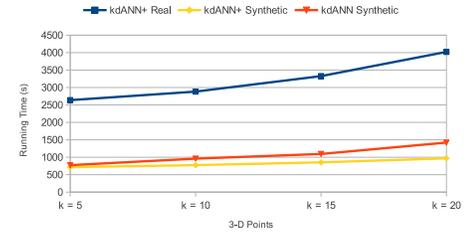,scale=.39}
\caption{Effect of $k$ for $d = 3$}
\label{fig10}
\end{figure}

\begin{figure*}
\centering
\subfloat[kdANN+]{
  \epsfig{file=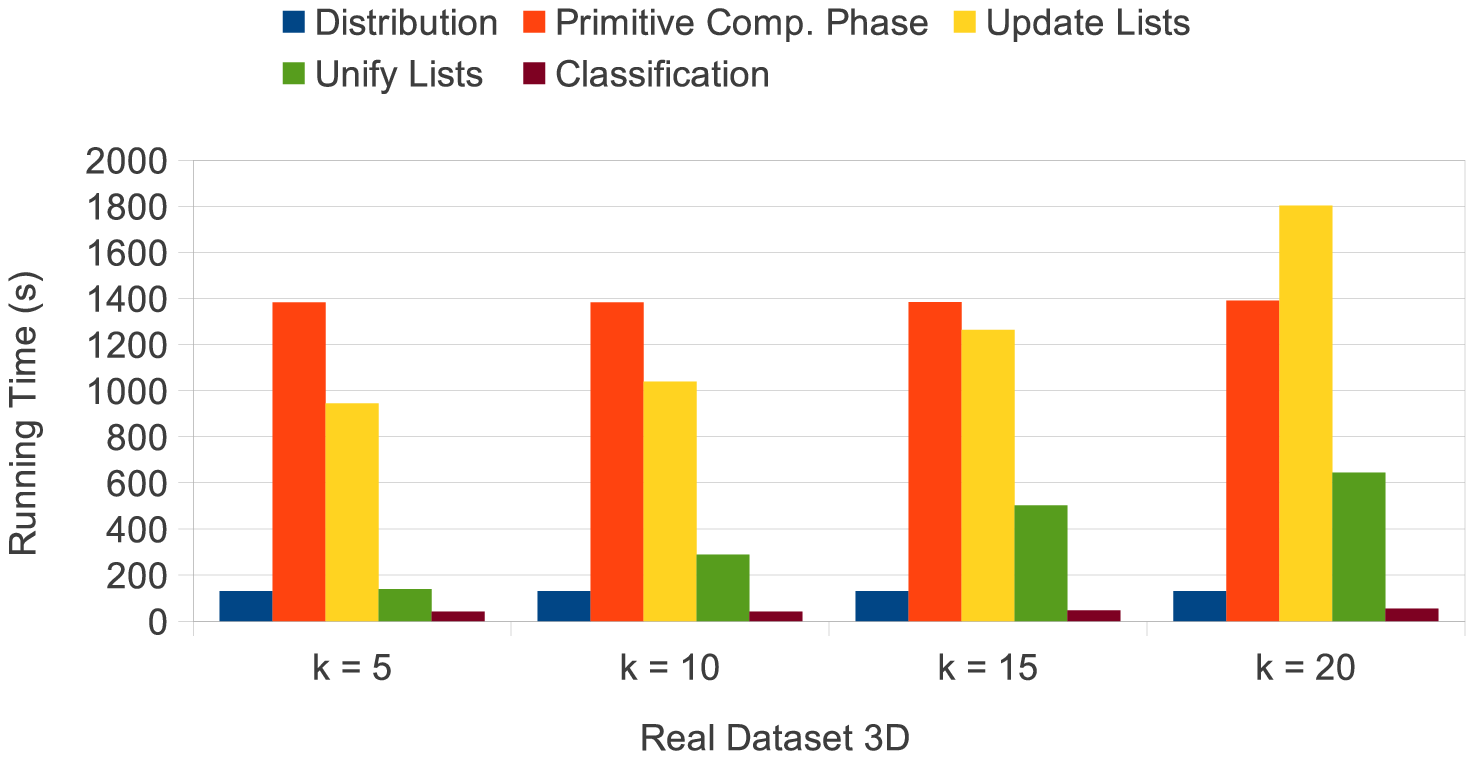,scale=.37,width=.3\linewidth}
  \label{fig:breakdown:a}
}
\qquad
\subfloat[kdANN+]{
  \epsfig{file=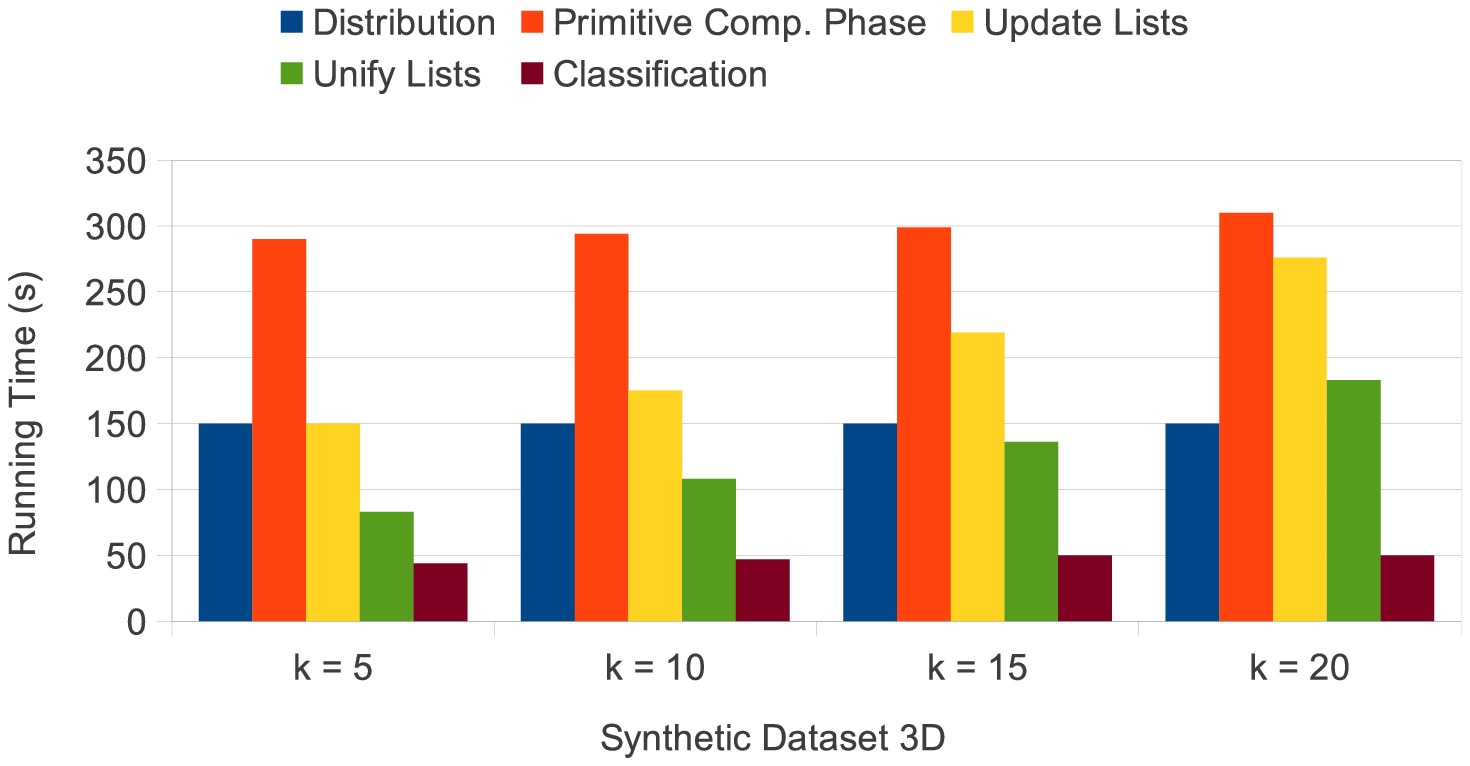,scale=.37,width=.3\linewidth}
  \label{fig:breakdown:b}
}
\quad
\subfloat[kdANN]{
  \epsfig{file=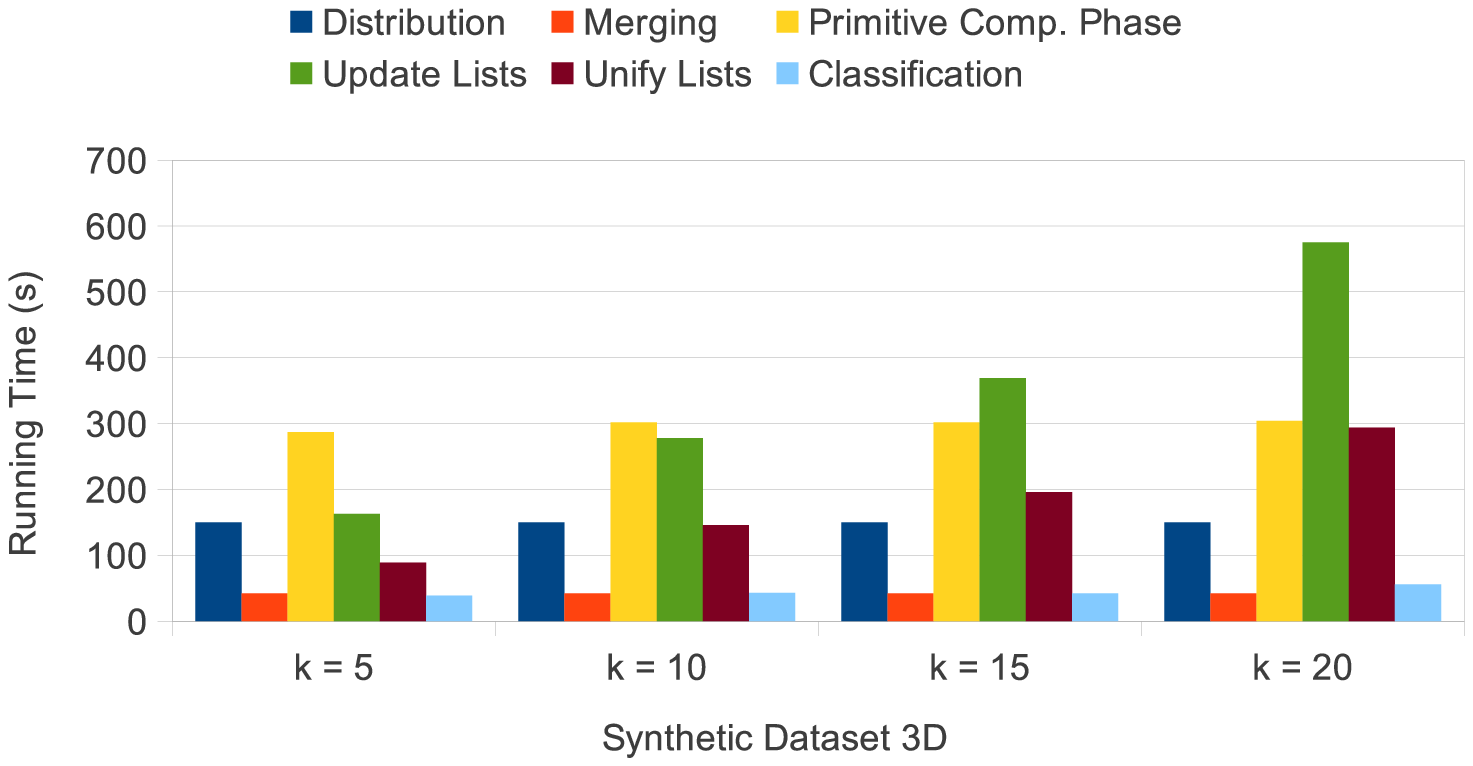,scale=.37,width=.3\linewidth}
  \label{fig:breakdown:c}
}
\caption{Phase breakdown vs $k$}
\label{fig:breakdown}
\end{figure*}

In Figure \ref{fig9}, we demonstrate the outcome of the experimental procedure for 2-dimensional points when we alter $k$ value from 5 to 20. First of all, note that we do not include the results of kdANN for the real dataset. The reason is that the method only produced results for $k = 5$ and needed more than 4 hours. Beyond this, the merging step of kdANN derived extremely sizeable cells and during the primitive computation phase a bottleneck was created to some nodes that strangled their resources, thus preventing them to derive any results. Observing the rest of the curves, we notice that the processing times are a bit higher than the previous ones due to larger records, as we impose one more dimension. Furthermore, the search area now overlaps more splits of the target space than in case of 1-dimensional points. Consequently, more instances of the $k$-NN lists are produced and we need more time to export the final ones. Overall, in the case of power law distribution, kdANN+ behaves much better than kdANN since the last one fails to process an A$k$NN query as $k$ increases. Also, kdANN+ is faster and in case of synthetic dataset that follows a uniform distribution, especially as $k$ increases.

Figure \ref{fig10} displays the results generated from kdANN and kdANN+ for the 3-dimensional points when we increase $k$ value from 5 to 20. Once again, in case of kdANN we could not get any results for any value of $k$ when we provided the real dataset as input. The reasons are the same we mentioned in the previous paragraph for $d = 2$. Table \ref{tab2} is pretty illustrative in the way the merging step affects the A$k$NN process. First of all, its computational cost is far from negligible if performed in a node (in contrary with the claim of the authors as stated in \cite{yoko:ishi:suzu}). Apart from this, the ratio of cubes that participate in the merging process is almost 40\% and the largest merged cube consists of 32,768 and 262,144 initial cubes for $k = 5$ and $k > 5$ respectively. In the case of kdANN+, when given the real dataset as input, it is obvious that the total computational cost is much larger compared to the one shown in Figures \ref{fig8} and \ref{fig9}. This happens for 3 reasons: 1) we have larger records in size, 2) some cubes are quite dense compared to others (since the dataset follows a power law distribution) and we need to perform more computations for them in the primitive computation phase and 3) a significant amount of overlaps occur, thus the update step of the $k$-NN lists needs more time than before. Finally, kdANN+ performs much better than kdANN, in the case of synthetic dataset, and the gap between the curves of running time tends to be bigger as $k$ increases.

In Figures \ref{fig:breakdown}\subref{fig:breakdown:a}-\ref{fig:breakdown}\subref{fig:breakdown:c} we present the results of running time for different stages of kdANN and kdANN+ as $k$ increases. We observe, that in all figures, the running time of distribution phase is the same (as it runs only once) while the running times of primitive computational and classification phase slightly increase. The running times of update and integrate phase increase notably, as the number of derived instances of the $k$-NN lists grows as $k$ increases. Finally, the running time of the merging phase remains the same. Apart from the merging phase, whose running time may increase significantly (Table \ref{tab2}), the running times for the rest phases follow the same trend as $d$ varies.

\begin{figure}
\centering
\epsfig{file=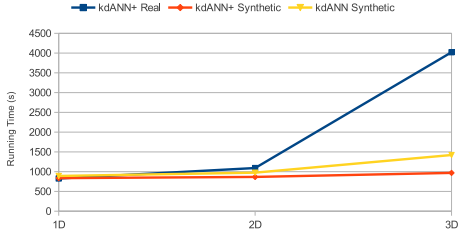,scale=.40}
\caption{Effect of dimensionality for $k = 20$}
\label{fig11}
\end{figure}

\begin{table}
\centering
\caption{Statistics of merging step for kdANN}
\label{tab2}
\resizebox{\linewidth}{!}{\begin{tabular}{|l|l|l|l|l|} \hline
 $ $ & $k = 5$ & $k = 10$ & $k = 15$ & $k = 20$ \\ \hline
 Time (s) & 271 & 675 & 962 & 1,528 \\ \hline
 \# of merged cubes & 798,032 & 859,944 & 866,808 & 870,784 \\ \hline
 \% of total cubes & 38\% & 41\% & 41.3\% & 41.5\% \\ \hline
 Max merged cubes & 32,768 & 262,144 & 262,144 & 262,144 \\ \hline
\end{tabular}}
\end{table}

\subsubsection{Effect of dimensionality}

In this subsection, we evaluate the effect of dimensionality for both real and synthetic datasets. Figure \ref{fig11} presents the running time for $k = 20$ by varying the number of dimensions from 1 to 3. 

From the outcome, we observe that kdANN is more sensitive to the number of dimensions than kdANN+ when we provide a dataset with uniform distribution as input. In particular, when the number of dimensions varies from 2 to 3 the divergence between the two curves starts growing faster. In the case of power law distribution, we only include the results for kdANN+ since kdANN fails to process the A$k$NN query for dimensions 2 and 3 when $k = 20$. We notice that the execution time increases exponentially when the number of dimensions varies from 2 to 3. This results from the curse of dimensionality. As the number of dimensions increases, the number of distance computations as well as the number of searches in neighboring ICCHs increases exponentially. Nevertheless, kdANN+ can still process the A$k$NN query in a reasonable amount of time in contrast to kdANN.

\subsubsection{Power Law vs Uniform Distribution}

In this subsection, we perform a comparative analysis of the results exported by our method for datasets with different distributions and argue about the performance of methods kdANN and kdANN+ as $k$ and $d$ increments. 

At first, we observe, that as $k$ increases kdANN+ prevails kdANN for all dimensions and for both dataset distributions. In case of uniform distribution, the divergence between the curves is not very big but the running time of kdANN+ increases linearly whilst kdANN's running time grows exponentially for $d > 2$. On the other hand, in case of power law distribution, for $d > 1$ kdANN+ outperforms kdANN, since the last one either fails to derive results in a reasonable amount of time or cannot produce any results at all. As shown in Table \ref{tab2}, the merging step has major deficiencies since it can cumber with notable computational burden the total A$k$NN process and can produce quite large merged ICCHs. As a consequence, the workload is badly distributed among the nodes and some of them end up running out of resources, thus causing kdANN to fail to produce any results. Finally, execution time of kdANN+ increases exponentially when the number of dimensions varies from 2 to 3.

Overall, the results exported by the experimental evaluation show that our solution (kdANN+) scales better, as $k$ and $d$ increases, than kdANN for uniform distributions and dominates it for power law distributions.

\subsection{Scalability}

In this experiment, we investigate the scalability of the two approaches. We utilize the 3$D$ datasets, since their size is bigger than the others, and create new chunks smaller in size that are a fraction $F$ of the original datasets, where $F \in $ \{0.2, 0.4, 0.6, 0.8\}. Moreover, we set the value of $k$ to 5.

\begin{figure}
\centering
\epsfig{file=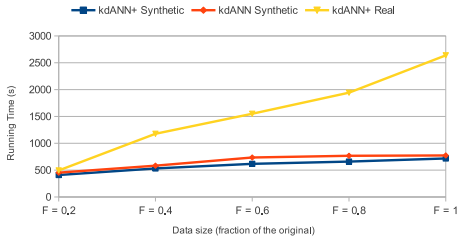,scale=.40}
\caption{Scalability}
\label{fig12}
\end{figure}

Figure \ref{fig12} presents the scalability results for real and synthetic datasets. In the case of power law distribution, the results display that kdANN+ scales almost linearly as the data size increases. In contrast, kdANN fails to generate any results even for very small datasets since the merging step continues to be an inhibitor factor in kdANN's performance. In addition, we can see that kdANN+ scales better than kdANN in the case of synthetic dataset and the running time increases almost linearly as in the case of power law distribution. Regarding kdANN, the curve of execution time is steeper until $F = 0.6$ and after that it increases more smoothly.

\begin{table}[t]
\centering
\caption{Statistics of merging step for kdANN and different data sizes}
\label{tab3}
\resizebox{\linewidth}{!}{\begin{tabular}{|l|l|l|l|l|} \hline
 $ $ & $F = 0.2$ & $F = 0.4$ & $F = 0.6$ & $F = 0.8$ \\ \hline
 Time (s) & 598 & 223 & 279 & 300 \\ \hline
 \# of merged cubes & 825,264 & 767,768 & 768,256 & 802,216 \\ \hline
 \% of total cubes & 39.3\% & 36.6\% & 36.6\% & 38.2\% \\ \hline
 Max merged cubes & 32,768 & 32,768 & 32,768 & 32,768 \\ \hline
\end{tabular}}
\end{table}

Table \ref{tab3} shows the way the merging step affects kdANN as the data size varies. The ratio of cubes that are involved in the merging process remains high and varies from 36.6\% to 39.3\% and the largest merged cube comprises of 32768 cubes of the initial space decomposition. Interestingly, the time to perform the merging step is not strictly increasing proportionally to the data size. In particular, the worst time is achieved when $F = 0.2$, then it reaches its minimum value for $F = 0.4$ and beyond this value augments again. Below, we explain why this phenomenon appears. The merging process takes into account the distribution information both of $I$ and $T$. As the size of the input dataset decreases, respectively the size of the training dataset also mitigates. Since both datasets follow a power law distribution, the ICCHs that include training set points decrease also in number and this may result in more merging steps (i.e. $F = 0.2$). 

\subsection{Speedup}

In our last experiment, we measure the effect of the number of computing nodes. We test four different cluster configurations and the cluster consist of $N \in \{11, 18, 25, 32\}$ nodes each time. We test the cluster configurations against the 3-dimensional datasets when $k = 5$.

\begin{figure}
\centering
\epsfig{file=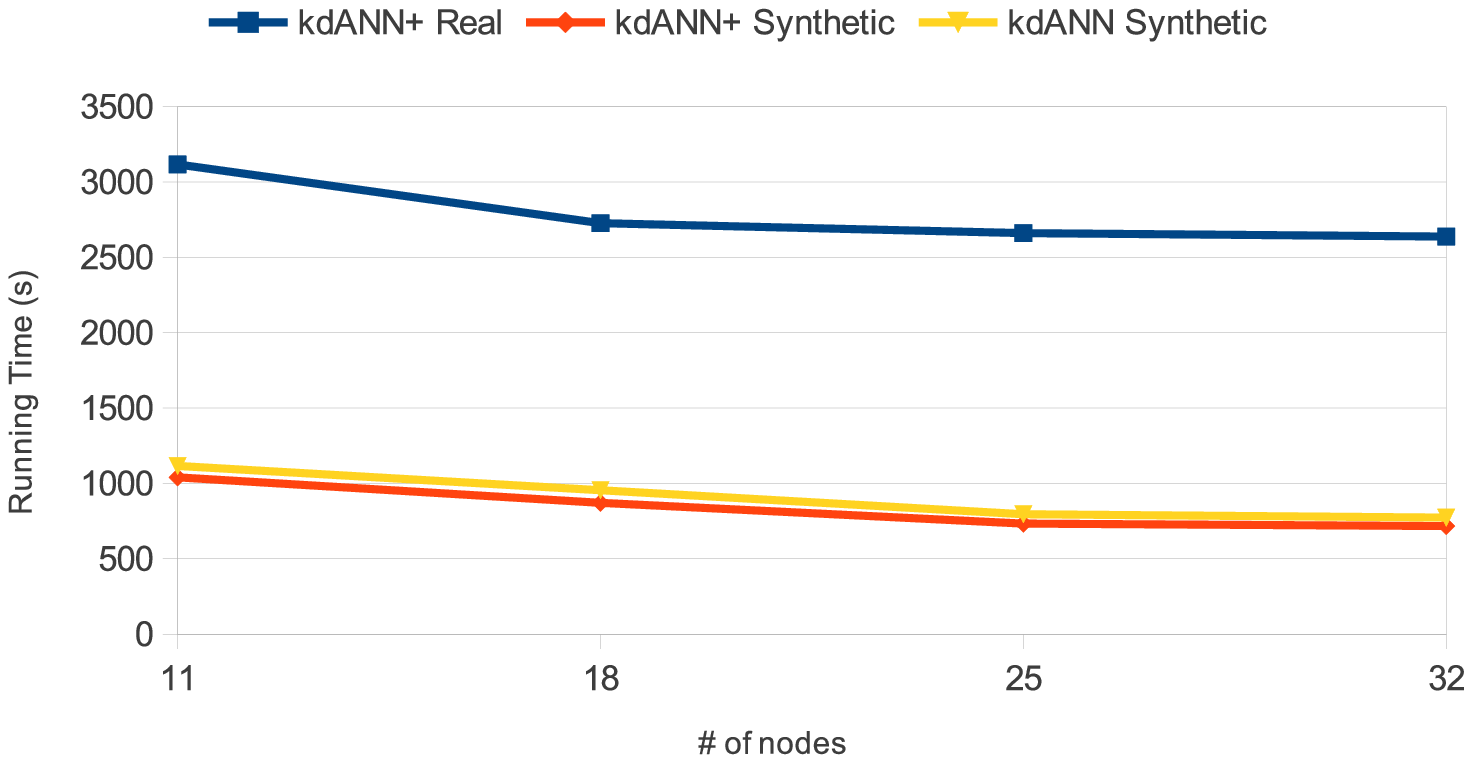,scale=.40}
\caption{Speedup}
\label{fig13}
\end{figure}

From Figure \ref{fig13}, we observe that total running time of kdANN+, in the case of power law distribution, tends to decrease as we add more nodes to the cluster. Due to the increment of number of computing nodes, the amount of distance calculations and update steps on $k$-NN lists that undertakes each node decreases respectively. Moreover, since kdANN fails to produce any results using 3-dimensional real dataset when the cluster consists of 32 nodes, it is obvious that it will fail with less nodes too. That is the reason for the absence of kdANN's curve from Figure \ref{fig13}. In the case of synthetic dataset, we observe that both kdANN and kdANN+ achieve almost the same speedup as the number of nodes increases; still kdANN+ performs betters than kdANN. We behold that in the case of real dataset the curve of running time decreases steeper as the number of nodes varies from 11 to 18 and becomes smoother beyond this point. On the other hand, in case of synthetic dataset the curves decrease smoother when the number of nodes varies from 25 to 32. The conclusion that accrues from this observation is that the increment of computing codes has a greater effect on the running time of both approaches when the datasets follow a uniform distribution. This happens because the workload is distributed better among the nodes of the cluster.

\subsection{Classification Performance}
In this section, we present the performance results of our classification method for kdANN+, when the 3$D$ real dataset is provided as input and $k = 10$. We define a set $C_T = \{A, B, C, D, E\}$ of 5 classes over the target space, but only 3 of them ($A, B, C$) contain points of $T$. The class where a point $t \in T$ belongs, depends on its coordinate vector. In Table \ref{tab4}, we measure the classification performance using four metrics for each class, \textit{True Positive, False Negative, False Positive} and \textit{True Negative} and give an average on the performance of each metric for all the classes. Among the classes, class $C$ has the worst accuracy but the overall results show that our classification method performs well.

\begin{table}
\centering
\caption{Classification performance of kdANN+}
\label{tab4}
\resizebox{\linewidth}{!}{\begin{tabular}{|l|l|l|l|l|} \hline
 $ $ & Class $A$ & Class $B$ & Class $C$ & Average \\ \hline
 True Positive & 99.97\% & 99.91\% & 94.14\% & 98\% \\ \hline
 False Negative & 0.03\% & 0.09\% & 5.86\% & 2\%\\ \hline
 False Positive & 0.13\% & 0.06\% & 0.06\% & 0.08\%\\ \hline
 True Negative & 99.87\% & 99.94\% & 99.94\% & 99.92\%\\ \hline
\end{tabular}}
\end{table}

\section{Conclusions}
In the context of this work, we presented a novel method for classifying multidimensional data using A$k$NN queries in a single batch-based process in Hadoop. To our knowledge, it is the first time a MapReduce approach for classifying multidimensional data is discussed. By exploiting equal-sized space decomposition techniques we bound the number of distance calculations we need to perform for each point to reckon its $k$-nearest neighbors. We conduct a variety of experiments to test the efficiency of our method on both, real and synthetic datasets. Through this extensive experimental evaluation we prove that our system is efficient, robust and scalable.

\section{Future Work}

\balance
In the near future, we plan to extend and improve our system in order to become more efficient and flexible. At first, we want to relax the condition of decomposing the target space into equal-sized splits. We have in mind to implement a technique that will allow us to have unequal splits that will contain approximately the same number of points. This is going to decrease the number of overlaps and calculations for candidate $k$-NN points. Moreover, in this way our method will be distribution independent and the load balancing between the nodes will be almost equal. 

In addition, we intend to apply a mechanism in order for the cluster to be used in a more elastic way, by adding (respectively removing) nodes as the number of dimensions increase (respectively decrease) or the data distribution becomes more (respectively less) challenging to handle.

Finally, we plan to use indexes, such as R-trees or M-trees, along with HBase, in order to prune any points that are redundant and cumber additional cost to the method.

\section{Acknowledgments}
This work was partially supported by Thales Project entitled \enquote{Cloud9: A multidisciplinary, holistic approach to internet-scale cloud computing}. For more details see the following URL:

\noindent \url{https://sites.google.com/site/thaliscloud9/home}

\bibliographystyle{abbrv}
\bibliography{aknn_classification}

\end{document}